\documentclass[useAMS,usenatbib]{mnras}
\usepackage{graphicx}
\DeclareGraphicsExtensions{.png,.pdf,.jpg}
\usepackage{float}
\restylefloat{figure}
\usepackage{caption}
\usepackage{subcaption}
\usepackage[authoryear]{natbib}
\usepackage{color}
\usepackage{amsmath}
\usepackage{amssymb}	% Extra maths symbols

\usepackage{ulem}

\title[The bilateral structure of the SNR G296.5+10]{Origin of the bilateral structure of the supernova remnant G296.5+10}

\author[A. Moranchel-Basurto et al.]{A. Moranchel-Basurto $^{1,2}$\thanks{E-mail:amoranchelb0500@alumno.ipn.mx},
P.~F. Vel\'azquez $^{2}$, E. Giacani $^{3,4}$, J.~C. Toledo-Roy $^{5}$,  \and  E.~M. Schneiter $^{6,7}$, F. De Colle $^{2}$, and A. Esquivel $^{2}$.
\\
$^{1}$Escuela Superior de F\'{\i}sica y Matem\'aticas, Instituto Polit\'ecnico Nacional,  CP: 07738, Mexico City, M\'exico\\
$^{2}$Instituto de Ciencias Nucleares, Universidad Nacional Aut\'onoma de M\'exico, CP: 04510, Mexico City, M\'exico\\
$^{3}$ Universidad de Buenos Aires, Facultad de Arquitectura, Dise\~no y Urbanismo, Departamento de Dise\~no Industrial, CP: 1428, \\ Buenos Aires, Argentina\\
$^{4}$ CONICET-Universidad de Buenos Aires, Instituto de Astronom\'\i a y F\'\i sica del Espacio (IAFE), CP: 1428, Buenos Aires, Argentina\\
$^{5}$Centro de Ciencias de la Complejidad, Universidad Nacional Aut\'onoma de M\'exico, CP: 0410, Mexico City, M\'exico.\\
$^{6}$Instituto de Astronom\'\i a  Te\'orica y Experimental, CP: X5000BGR, C\'ordoba,  Argentina\\
$^{7}$Departamento de Materiales y Tecnolog\'\i a, Universidad Nacional de C\'ordoba, CP: X5016GCA, C\'ordoba, Argentina.
}

\date{Accepted XXX. Received YYY; in original form ZZZ}

\pubyear{2017}
\begin{document}

\label{firstpage}
\pagerange{\pageref{firstpage}--\pageref{lastpage}}
\maketitle

\begin{abstract}
In the present work we have modeled the supernova remnant (SNR) G296.5+10, by means of
3D magnetohydrodynamics (MHD) simulations. This remnant belongs to the bilateral SNR group and has an additional striking feature: the rotation measure (RM) in its eastern and western parts are very different.
In order to explain both the morphology observed in radio-continuum and the RM, we consider that the remnant expands into a medium shaped by the superposition of the magnetic field of the progenitor star with a constant Galactic magnetic field.
We have also carried out a polarization study from our MHD results, obtaining synthetic
maps of the linearly polarized intensity and the Stokes parameters. This study reveals that both the radio morphology and the reported RM for G$296.5+10$ can be explained if the quasi-parallel acceleration mechanism is taking place in the shock front of this remnant.
\end{abstract}
%%%%%%%%%%%%%%%%%%%%%%%%%%%%%%%%%%%%%%%%%%%%%%%%%%

%%%%%%%%%%%%%%%%% BODY OF PAPER %%%%%%%%%%%%%%%%%%

\section{Introduction}

Recently, the study of bilateral supernova remnants \citep[BSNRs, ][]{gaensler1998},
also called `barrel-shaped' \citep{kesteven1987} or `bipolar' \citep{fulbright1990}, has
gained great interest since they have proven to be a very
useful tool when studying the configuration of the interstellar
magnetic field (ISMF) within a few parsecs around the SNR \citep{bocchino2011,reynoso2013,west2016}. As suggested by their name, the BSNRs are characterized by a clear axis of symmetry, given by two bright arcs separated by a region of low surface brightness.

%In general, the remnants appear asymmetric, distorted and
%elongated with respect to the shape and surface brightness
%of the two opposed arcs.

In spite of the increasing interest in BSNRs, a satisfactory and complete model that explains
the observed morphology and the origin of the asymmetries does not yet exist.
Attempts to explain this morphology have focused on
two models: the equatorial belt model (quasi-perpendicular mechanism of acceleration)
and the polar cap model (quasi-parallel mechanism). In the first model, the orientation of the ISMF is perpendicular to the shock front normal, while in the second model the orientation of the ISFM is parallel
to the shock front normal \citep{fulbright1990}.

The SN G$296.5+10$ remnant, also known as PKS 1209-51/52, is a member of the BSNR group. The radio-continuum and the X-ray emissions present two opposite bright arcs with their symmetry axis
oriented almost perpendicular to the Galactic plane.
This remnant has a radio size of $90'\times65'$, and it is located at a distance of $2.1$ kpc \citep{Giacani2000}.

The bilateral radio morphology of G$296.5+10$ has been explained by \citet{orlando2007} as a product of the expansion of the SNR into an interstellar medium (ISM) with a gradient in either magnetic field or density. They presented synthetic synchrotron maps considering both acceleration mechanisms \citep[see figures 6 and 7 of][]{orlando2007} and concluded that the quasi-perpendicular mechanism is the dominant.

Recently, \citet{west2016} carried out a survey of Galactic bilateral SNRs that included the SNR G296.5+10. They modeled the linearly polarized intensity and also came up to the conclusion that the morphology of this remnant can be explained by considering the quasi-perpendicular acceleration mechanism.

The rotation measure (RM) of G$296.5+10$ has a striking and interesting feature that could help constrain models.
Its eastern and western parts show large differences, as reported by \citet{Harvey-Smith2010} \citep[see also][]{Whiteoak-gardner68,dickel76}, where the RM even shows a change in sign. \citet{Harvey-Smith2010} suggested that this asymmetry can be explained with an azimuthal magnetic field in the stellar wind of a red supergiant (RSG) progenitor.

The aim of this paper is to find out, following the idea of \citet{Harvey-Smith2010}, whether the history of the progenitor star could explain the observed RM and have any influence on the morphology of SNR G296.5+10. Thus, we have carried out MHD simulations modeling a supernova remnant which  expands into a medium which has been previously swept up by the wind of a RSG. For the magnetic field we have considered two components: (1) the magnetic field of the RSG star, and (2) a uniform component associated with the Galactic magnetic field. From our MHD results,
we carried out a polarization study in order to determine which acceleration
mechanism of relativistic particles achieves a better agreement with observations.

    The present paper is structured as follows: section 2
describes the model and the techniques used to perform the
simulations. In section 3 we explain how synthetic radio maps were obtained from numerical results. Section 4 present the results, and finally in section 5 we discuss the results and make the final remarks.

\section{Numerical Model}

\subsection{Code description}
\label{sec:model} % used for referring to this section from elsewhere
The numerical simulations were carried out with the parallel 3D MHD code {\sc Mezcal} \citep{DeColle&Raga05,DeColle&Raga06}. In this code the shock propagation is modeled by numerically solving the time-dependent ideal MHD equations
of mass (eq. \ref{eq:mass}), momentum and energy conservation (eq. \ref{eq:momentum} and \ref{eq:energy}) together with the induction equation (eq. \ref{eq:induction}) and a rate equation (eq. \ref{eq:hydrogen}) for the ionization of hydrogen. We employ a 3D
Cartesian coordinate system with a binary adaptive mesh, and consider the cooling function $\Lambda$
of \citet{DeColle&Raga05} \citep[see also,][]{DeColle&Raga06}.
The aforementioned equations are:

\begin{equation}
    \frac{\partial\rho}{\partial t}+\nabla\cdot(\rho\mathbf{v})=0\;,
	\label{eq:mass}
\end{equation}

\begin{equation}
    \frac{\partial\rho\mathbf{v}}{\partial t}+\nabla\cdot(\rho\mathbf{v}\mathbf{v}+p_{\text{tot}}\mathbf{I}-\mathrm{\mathbf{B}}\mathrm{\mathbf{B}})=0\;,
	\label{eq:momentum}
\end{equation}

\begin{equation}
    \frac{\partial e}{\partial t}+\nabla\cdot((e+p_{\text{tot}})\mathbf{v}-(\mathbf{v}\cdot\mathrm{\mathbf{B}})\mathrm{\mathbf{B}})=-n^2\Lambda\;,
	\label{eq:energy}
\end{equation}

\begin{equation}
    \frac{\partial \mathrm{\mathbf{B}}}{\partial t}+\nabla\cdot(\mathbf{v}\mathrm{\mathbf{B}} - \mathrm{\mathbf{B}}\mathbf{v})=0\;,
	\label{eq:induction}
\end{equation}

\begin{equation}
    \frac{\partial n_{\mathrm{H}^0}}{\partial t}+\nabla\cdot(n_{\mathrm{H}^0}\mathbf{v})=n_{\mathrm{H}^{+}}n_{\mathrm{e}}\alpha(T)-n_{\mathrm{H}^0} n_{\mathrm{e}} C(T)\;,
	\label{eq:hydrogen}
\end{equation}

\noindent where $\rho$ is the mass density, $\mathbf{v}$ is the velocity vector field, $p_{\text{tot}}=p_{\text{gas}}+B^2/2$ is the (magnetic+thermal) total pressure, $\mathbf{I}$ is the identity matrix, $\mathrm{\mathbf{B}}$ is the magnetic field normalized by  $\sqrt{4\pi}$, $e$ is the total energy density defined as $e=p_{\text{gas}}/(\gamma-1)+\rho v^2/2+\rho B^2/2$ (where $\gamma=5/3$ is the adiabatic index), $\Lambda(T)$ is the cooling function, and $T$ is the temperature. In the last equation, $n_{\mathrm{H}^0}$, $n_{\mathrm{H}^+}$, and $n_{\mathrm{e}}$ are the  neutral hydrogen, ionized hydrogen, and electron number densities, respectively. $\alpha(T)$ and $C(T)$ are the recombination and collisional ionization coefficients.

\subsection{Effects of rotation and magnetic fields in the wind of the progenitor star}

Previous works have explored the evolution of a SNR in a cavity generated by the progenitor stellar wind. For example  \citet{Dwarkadas07b} \citep[see also][]{Dwarkadas} studied the formation of the stellar bubble following the evolution of the progenitor, starting as an O-type star, passing through the Super Red Giant phase and ending as a Wolf-Rayet star. This study was carried out by means of 1D and 2D hydrodynamical simulations, considering the changes in the mass loss rates and terminal velocity of the stellar wind in each phase of evolution. These stellar winds generate different shells, which after a time collide with each other. This interaction produces a turbulent cavity, where the SNR ultimately evolves.
In the present work we have carried out a 3D MHD simulation and, as a first approach to this problem, we evolve the SNR in a medium which has been swept up by the progenitor stellar wind, with an initial density profile given by the free wind solution and spanning the whole computational domain. This density distribution is imposed before the passage of the SNR shock wave.

Then, we consider a spherical stellar wind with a constant mass-loss rate $\dot{M}$ and an asymptotic velocity {$V_{\text{sw}}$}. Mass conservation implies that the number density $n(r)$
of the stellar wind cavity at a radius $r$ from the center of the star (where $r$ lies
outside the wind acceleration region) is given by:

\begin{equation}
    n(r)=\frac{\dot{M}}{4\pi r^2V_{sw}m_{\mathrm{H}} }\;,
\label{n_wind}
\end{equation}

\noindent where $m_\mathrm{H}$ is the mass of a hydrogen atom.

For a rotating magnetized star, the wind is channeled by the magnetic field lines, since the plasma
is frozen-in to the field. Magnetic field lines leave the surface
of the star radially and rapidly wind up into a spiral shape
at radii greater than the maximum co-rotation radius. For a star of radius $r_c$ with an equatorial rotational velocity $v_{\text{rot}}$, and assuming symmetry above and below the equatorial
plane, the strength of the azimuthal component of the stellar magnetic field $B_{\varphi}(r)$, for $r\gg r_c$, \citep{garcia1999} is\footnote{The radial component is negligible with respect to the azimuthal component for $r \gg r_c$.}:
\begin{equation}
B_{\varphi}(r)=B_{r}(r_c)\frac{r_c}{r}\frac{v_{rot}}{V_{sw}}\cos(\delta)\;,
\label{B_tor}
\end{equation}
where $\delta$ is the latitude (measured with respect to the equator, the $xz$-plane), and $B_r(r_c)$ is the magnetic field strength on the star surface ($r=r_c$).

Additionally, we considered a constant component for the Galactic magnetic field $\mathrm{\mathbf{B}_G}$. Thus, the
total initial magnetic field $\mathrm{\mathbf{B}_{0}}$ is the vectorial composition of $\mathrm{\mathbf{B}_G}$ and $\mathrm{\mathbf{B}_{\varphi}}$, i.e. $\mathrm{\mathbf{B}_{0}}=\mathrm{\mathbf{B}_G}+\mathrm{\mathbf{B}_{\varphi}}$.

\begin{figure}
  \centering
  \includegraphics[width=0.42\textwidth]{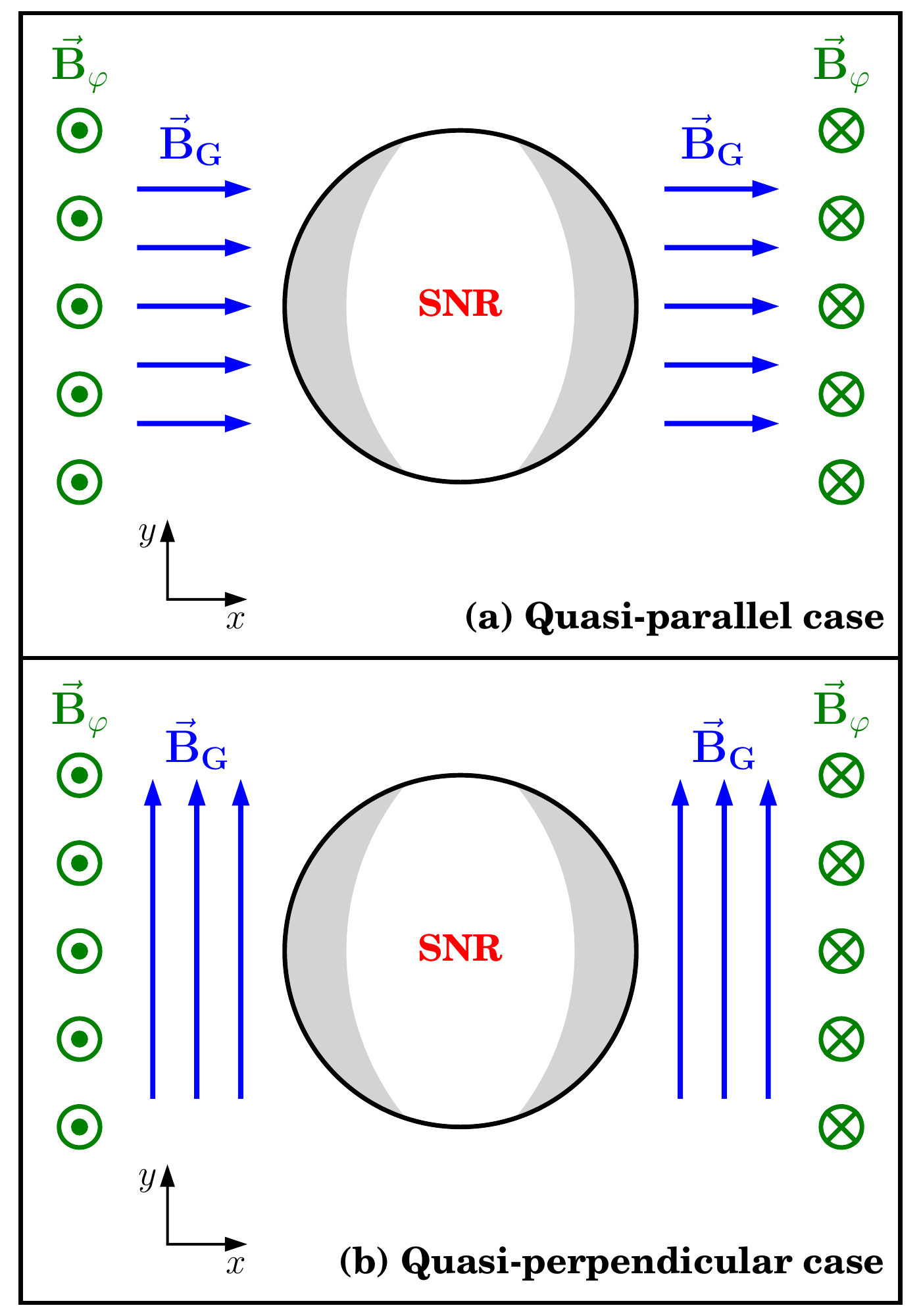}
  \caption{Diagram of the magnetic field configurations employed in our simulations. Blue arrows indicate the orientation of the Galactic magnetic field. The orientation of the azimuthal component of the progenitor magnetic field is indicated by the green vectors coming into and out of the page.}
  \label{fig:eff_cases}
\end{figure}

Two
configurations for $\mathrm{\mathbf{B}_G}$ were used depending on which particle acceleration process is assumed to be dominant. For the quasi-perpendicular case, $\mathrm{\mathbf{B}_G}$ is along the $y$-direction, while for the quasi-parallel case
it is directed along the $x$-direction (see Figure \ref{fig:eff_cases}).

As an ad hoc assumption, the magnitude of $\mathrm{\mathbf{B}_G}$ was chosen such that $B_G \simeq B_{\varphi}$  20 pc away from the center of the SNR (the physical size of the SNR)\footnote{In this way, both magnetic field components would have the same contribution to the synchrotron emission.},

Finally, we observe that the so-called ``quasi-perpendicular case'' would actually result in a total
field with a spiral-like pattern, while in the so-called ``quasi-parallel case'' the vectorial combination of azimuthal plus
uniform field would lead to a distorted azimuthal field, with a stronger total field to one side (doubling at 20 pc) and a weaker one
to the other side (vanishing at 20 pc).

\subsection{Initial Setup}

The supernova explosion was initialized by the deposition of an energy $E_0=5\times10^{50}$~erg (a typical value for a type II SN explosion) in a sphere of radius $r_0=1.3$ pc, containing a total mass of $2.5$~$\mathrm{M_{\odot}}$, located at the center of the computational domain. The remnant expands into the progenitor stellar wind, characterized by a mass-loss rate $\dot{M}=10^{-5}$ $\mathrm{M}_{\odot}~\text{yr}^{-1}$ and $V_{\text{sw}}=15$ $\text{km s}^{-1}$. The parameters of the azimuthal magnetic field (see Eq. \ref{B_tor}) were $r_c=6.8\times10^{-6}$~pc, $v_{rot}=0.07~V_{\text{sw}}$, and $B_{r}(r_c)=300$~G. The 3D computational domain is a cube with a physical size of 70 pc per side, with a resolution of $\sim 0.14\,\mathrm{pc}$ at the highest level of refinement. We follow the expansion of the remnant until an integration time of $8000$~yr.

\subsection{The turbulent background}
In order to simulate the expansion of the SNR in a more realistic environment, we introduced turbulent-like perturbations in the magnetic field.
This perturbation has a 3D power spectrum that is consistent with a Kolmogorov type cascade, given by  \citep{jOKIPI1987, yu, Fang14}:
\begin{equation}
P\propto \frac{1}{1+(kL_c)^{11/3}}\;.
\end{equation}
A coherence length of $L_c=3$~pc was assumed, with the wavenumber $k=\frac{2\pi}{L}$ obtained by varying $L$ from $\Delta x$ (the size of a computational cell) to $L_{\text{sim}}$ (the size of the computational domain). Also, $N_m=900$ wave modes were considered in order to simulate an isotropic turbulent medium.

The initial magnetic field configuration for each point of the computational domain $(x,y,z)$ is given by:
\begin{equation}
{\bf B}(x,y,z)={\bf B}_0(x,y,z)+\delta {\bf B}(x,y,z)\;,
\end{equation}
where ${\bf B}_0(x,y,z)$ is the total unperturbed magnetic field. The magnetic field perturbation $\delta {\bf B}(x,y,z)$ is given by \citep{yu,velazquez2017}:

\begin{equation}
\delta \mathrm{\mathbf{B}}(x,y,z) = \Re \left[\sum_{n=1}^{N_m} A(k_n)(\cos\alpha_n\hat{x}'+i\sin\alpha_n\hat{y}')e^{ik_nz_n'} \right]\;,
\label{deltaB}
\end{equation}
where
\[
\begin{pmatrix}
 x' \\
 y' \\
 z'
\end{pmatrix}
=
\begin{pmatrix}
%\[ \left( \begin{array}{ccc}
\cos \theta_n \cos \phi_n& \cos \theta_n \sin \phi_n & -\sin \theta_n \\
-\sin \phi_n & \cos \phi_n &  0 \\
\sin \theta_n \cos \phi_n & \sin \theta_n \sin \phi_n & \cos \theta_n
\end{pmatrix}
\begin{pmatrix}
 x \\
 y \\
 z
\end{pmatrix}
\]
where $\theta_n$ and $\phi_n$
represent  the  direction  of  propagation  of  the  wave
mode $n$ which a wave number $k_n$ and  polarization $\alpha_n$.

The amplitude $A(k_n)$ in  Eq.(\ref{deltaB}) is given by:
 \begin{equation}
A(k_n)=\sigma_B^2\frac{\Delta V_n}{1+(k_nL_c)^{11/3}}\sum_{n=1}^{N_m}\left[\frac{\Delta V_n}{1+(k_nL_c)^{11/3}}\right]^{-1}\;,
\end{equation}
where $\sigma_B^2$ is the wave variance of the magnetic field ($\sigma_B^2=(0.25 B_G)^2$ in this work), and the normalization factor $\Delta V_n$ is given by $\Delta V_n=4\pi k_n^2\Delta k_n$. The wave numbers $k_n$ are spaced logarithmically so that $\Delta k_n/k_n$ remains constant \citep{giacalone-jokipi}.

% Both systems $(x,y,z)$ and $(x',y',z')$ are related by:
% \begin{equation}
%  \begin{pmatrix} x' \\ y' \\ z' \end{pmatrix} =\begin{pmatrix} cos\theta_ncos\phi_n & cos\theta_nsen\phi_n & -sen\theta_n \\ -sen\phi_n & cos\phi_n & 0 \\ sen\theta_ncos\phi_n & sen\theta_nsen\phi_n & cos\theta_n \end{pmatrix}\begin{pmatrix} x \\ y \\ z \end{pmatrix}
% \end{equation}

% For the density fluctuation we employed the same log-normal distribution as in \cite{giacalone-jokipi}:
% \begin{equation}
% n(x,y,z)=n_0e^{f_0+\delta f}
% \end{equation}
% where $f_0$ is a constant and $\delta f$ is the density perturbation with a wave variance $\sigma_d^2$. Both $\delta f$ and $\delta \vec{B}$ have the same 3D Kolmogorov-like power spectrum.
% The turbulent environment temperature and number density were set to $T_0=10^4K$ and $n_0=5\times 10^2$, respectively. The wave variances for the magnetic field and density were set as $\sigma_B^2=(0.25B_0)^2$.

\section{Synthetic Maps}

\begin{figure*}
   \centering
   \includegraphics[width=0.85\textwidth]{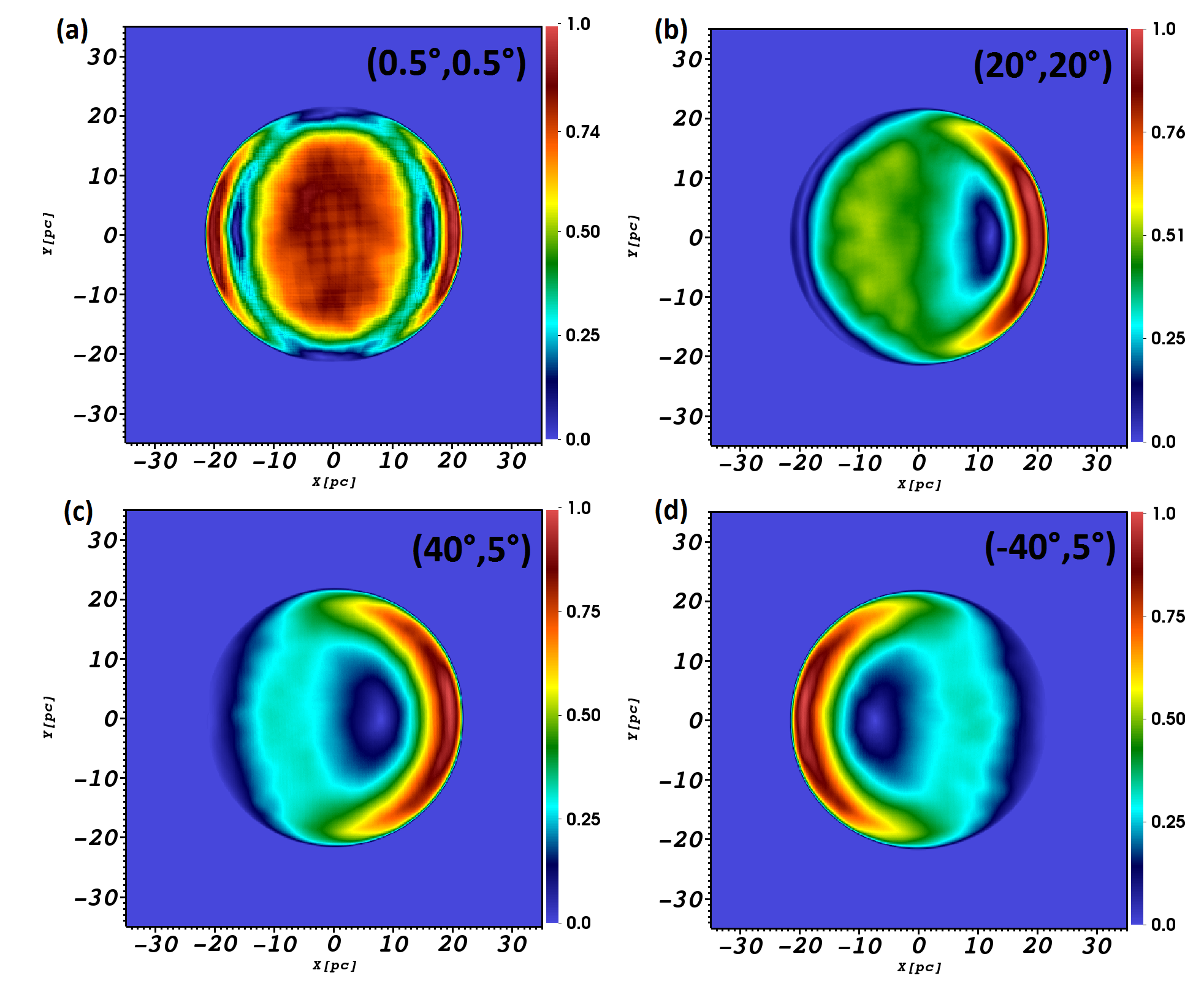}
   \caption{Synthetic maps of the linearly polarized emission, at different rotation angles (with respect to the $x-$ and $y-$ axes, which are indicated on the top right corner of each panel, in units of degrees), obtained for the quasi-perpendicular case. The color scale gives the normalized flux. Both axis are given in units of pc.}
   \label{fig:sinc_bper}
\end{figure*}

\subsection{Synchrotron emission}
We estimated the radio emission from the remnant, assuming that it is
due to synchrotron radiation from relativistic electrons, with a power law spectrum distribution $N(E)=KE^{-\xi}$ where $E$ is the electron energy, $N(E)$ is the number of electrons per unit volume with arbitrary directions of motion
and with energies in the interval $[E, E + dE]$, $K$ is the normalization
of the electron distribution, and $\xi$ is the power law index. Following \citet{JUN96...465..800J} (and references therein), the radio
emissivity is calculated for each cell in the computational domain $(x_i,y_i,z_i)$ as:

\begin{equation}
    i(x_i,y_i,z_i,\nu)=C_1\ K\ p^{2\alpha} \rho^{1-2\alpha} B_{\perp}^{\alpha+1}\nu^{-\alpha}\;,
    \label{isynchro}
\end{equation}

\noindent where $C_1$ is a constant which is assumed to be 1\footnote{In this work we are interested in carrying out a qualitative analysis of the synchrotron emission, i.e. we do not expect to reproduce the full quantitative synchrotron flux of this object.}, $p$ is the gas pressure, $\rho$ is the density of the gas, $B_{\perp}$ is the component of the magnetic field perpendicular to the line-of-sight (LoS), $\nu$ is the frequency of the radiation, and $\alpha=(\xi-1)/2$ is the synchrotron spectral index.  We do not evolve the relativistic electron population, but rather assume a constant spectral index with a value of 0.5 \citep{milne1994}. Eq.(\ref{isynchro}) was obtained considering that the density and the energy of the relativistic particles are constant fractions of the density and thermal energy of the gas. On the other hand, in this work we have considered that the coefficient $K$ includes the dependence on the obliquity angle $\Theta_{Bs}$, being either proportional to $\sin^2(\Theta_{Bs})$, for the quasi-perpendicular case, or to $\cos^2{(\Theta_{Bs})}$, for the quasi-parallel case \citep{fulbright1990}. The angle $\Theta_{Bs}$ is the angle between the shock normal and the post-shock magnetic field.

\begin{figure*}
   \centering
   \includegraphics[width=0.85\textwidth]{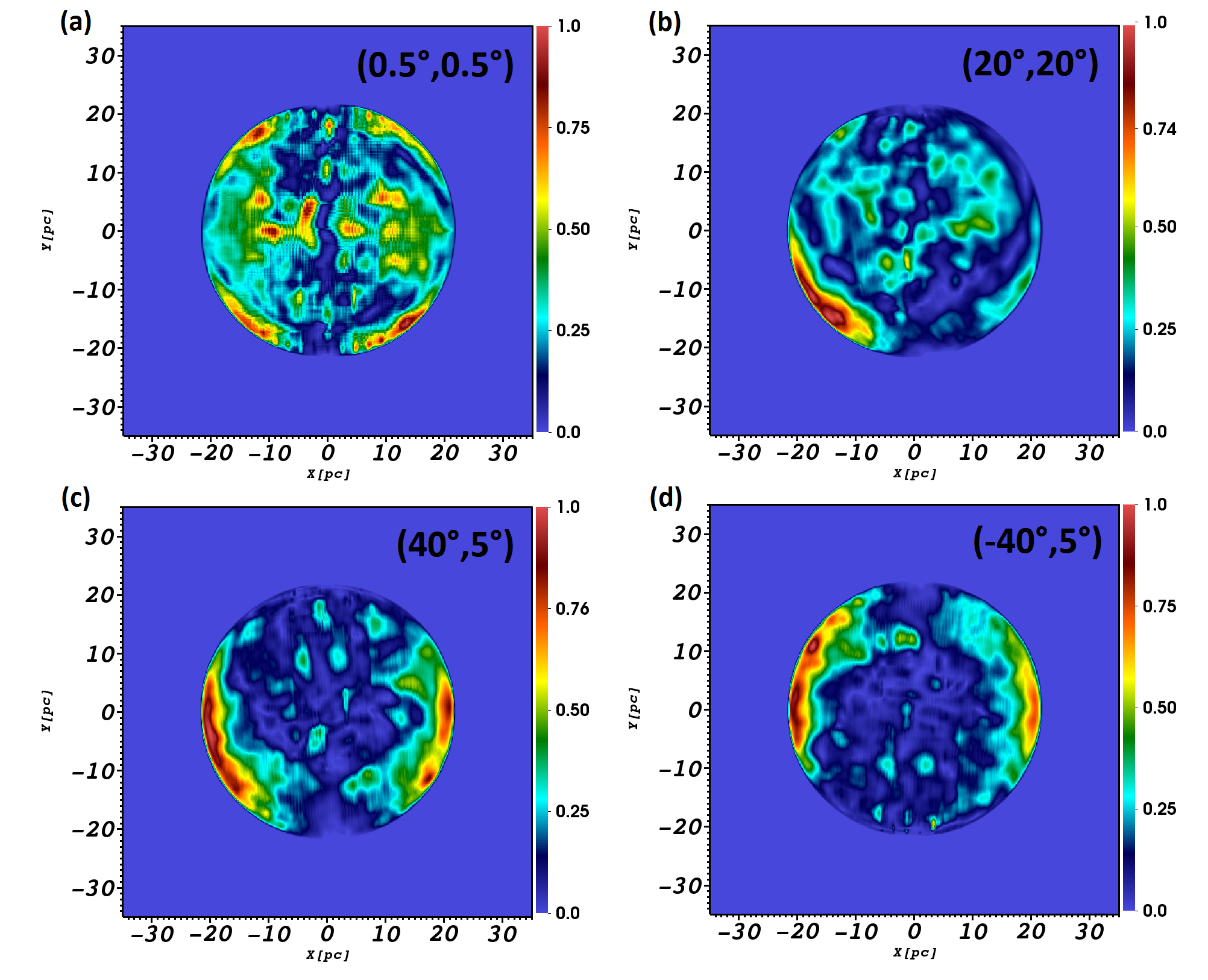}
   \caption{Same as Figure \ref{fig:sinc_bper} but for the quasi-parallel case.}
   \label{fig:sync_bpar}
\end{figure*}

\begin{figure}
   \centering
   \includegraphics[width=0.5\textwidth]{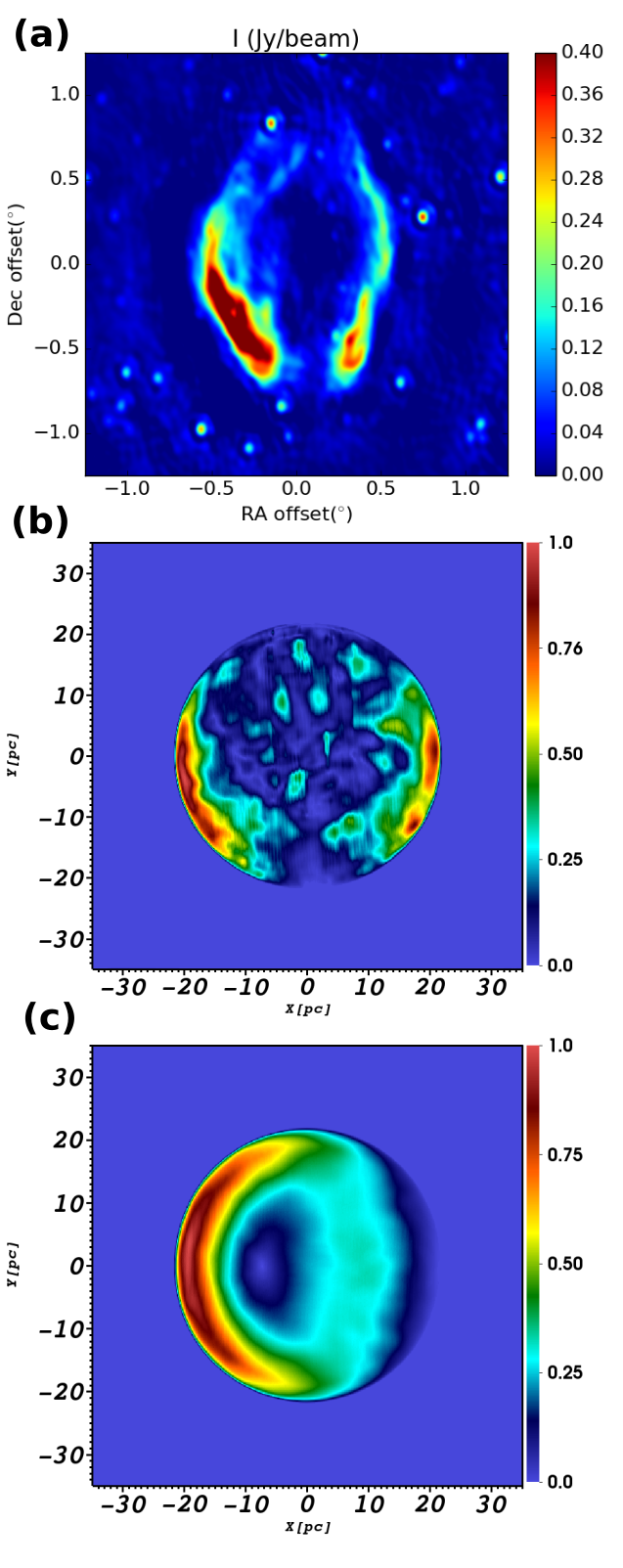}
   \caption{Comparison between the observed (panel {\it a}) and the synthetic maps of the linearly      polarized intensity  (panel {\it b} and {\it c}, corresponding to the quasi-parallel and the quasi-perpendicular cases, respectively). The axes are given in units of pc.}
   \label{fig:synchro}
\end{figure}

\subsection{Faraday Rotation}
When polarized electromagnetic radiation propagates through a magnetized ionized medium its electric field vector is rotated an angle which depends on the square of the wavelength and the path integral of the electron-weighted LoS magnetic field strength. This effect is called Faraday Rotation. The angle of rotation is $\Delta\chi=\mathrm{RM}~\lambda^2$, where the RM is defined as \citep{rohlfs}:

\begin{equation}
\mathrm{RM} = 0.81 \int_{\text{source}}^{\text{observer}} n_e B_{\parallel} dz_i \  \mathrm{rad\ m^{-2}}\;,
\label{RM}
\end{equation}
where $B_{\parallel}$ is the LoS component of the magnetic field strength in $\mathrm{\mu}$G, $n_e$ is the electron density in $\mathrm{cm^{-3}}$, and $dz_i$ is measured in pc.

\subsection{Stokes parameters}

To characterize the intensity and the state of polarization of the synchrotron radiation, we make use of the Stokes parameters. In this work, we obtain the synthetic maps for the Stokes Parameters Q and U by means of the following expressions \citep[see][]{clarke1989,jun1996b,Cecere2016}:

\begin{equation}
Q(x_i,y_i,\nu)=\int\limits_{\mathrm{LoS}}f_0i(x_i,y_i,z_i,\nu)\cos(2\phi(x_i,y_i,z_i))dz_i\;,
\label{stokesQ}
\end{equation}

\begin{equation}
U(x_i,y_i,\nu)=\int\limits_{\mathrm{LoS}}f_0i(x_i,y_i,z_i,\nu)\sin(2\phi(x_i,y_i,z_i))dz_i\;,
\label{stokesU}
\end{equation}
where $(x_i,y_i)$ are the coordinates on the plane of the sky (or image plane), and $z_i$ is the coordinate along the LoS, $\phi(x_i,y_i,z_i)$ is the position angle of the local electric field on the plane of the sky, and $f_0$ is the degree of linear polarization, which is a function of the spectral index $\alpha$:
\begin{equation}
f_0=\frac{\alpha+1}{\alpha+\frac{5}{3}}\;.
\label{fraccPol}
\end{equation}

The angle $\phi(x_i,y_i,z_i)$ is obtained from the position angle of the local magnetic field $\phi_{\mathrm B}(x_i,y_i,z_i)$, which is directly known from MHD simulations. To compare with the observational data, we follow the inverse process: the magnetic field was rotated clockwise $90^{\circ}$ and then we performed the inverse Faraday correction, i.e. the argument $\phi(s)$ in the eqs. (\ref{stokesQ}) and (\ref{stokesU}) was replaced by $\phi_{\mathrm B}(x_i,y_i,z_i)-\frac{\pi}{2}+\Delta\chi_F$
where
$\Delta\chi_F$ is the Faraday Rotation correction:
\begin{equation}
\Delta\chi_F = \frac{RM}{\text{1 rad m}^{-2}}\left(\frac{\lambda}{\text{1 m}}\right)^2\;,
\label{eq:Faraday}
\end{equation}
where $\lambda$ is the observed wavelength (in units of m).
For the case of SNR G$296.5+10$, \citet{Harvey-Smith2010} found that this remnant has a highly ordered RM structure with an average RM of $+28\ \mathrm{rad\ m}^{-2}$ on the eastern side, and a RM of $-14\ \mathrm{rad\ m}^{-2}$ on the western side. With these values and using  Eq. (\ref{eq:Faraday}) we obtained a correction angle due to Faraday Rotation of $1.32$~rad  or $75.4^{{\circ}}$ for the eastern part of the remnant (the left half of the remnant in the synthetic maps), and $-0.66$~rad or $-37.7^{\circ}$ for the western side (the right half in the synthetic maps), for an observed wavelength $\lambda$ of $0.21$~m.

Finally, the linearly polarized intensity and the position angle of the intrinsic magnetic field are given by the following expressions:
\begin{equation}
I_p(x_i,y_i,\nu)=\sqrt{Q(x_i,y_i,\nu)^2+U(x_i,y_i,\nu)^2},
\label{Ipol}
\end{equation}
and
\begin{equation}
\chi_B(x_i,y_i)=\frac{1}{2}\arctan \left(\frac{U_B(x_i,y_i,\nu)}{Q_B(x_i,y_i,\nu)} \right)
\label{chi_B}
\end{equation}
where $Q_B(x_i,y_i,\nu)$ and $U_B(x_i,y_i,\nu)$ were calculated by replacing $\phi(x_i,y_i,z_i)$ by $\phi_{\mathrm B}(x_i,y_i,z_i)$ in Eqs.(\ref{stokesQ}) and (\ref{stokesU}). The position angle $\chi_B$ (Eq. \ref{chi_B}) only gives the direction of the magnetic field, it does not fully specify its orientation, i.e. the magnetic field in the plane of the sky is defined only up to $\pm \pi$.

\section{Results}

To compare our simulations with the observations, we reprocessed ATCA archival data of G$296.5+10$ at 1.4 GHz. These data were obtained during three sessions (of 13 hours each) on October 8-10, 1998, through a 109 pointings mosaic, and with them we obtained the I, Q, U, and V Stokes parameters. The source PKS B1934-638 was used for primary flux density, polarization leakage and bandpass calibration, while PKS B1215-457 was used to calibrate the phases. All data reduction and calibration was carried out using the {\sc miriad} package following standard procedures.

Figure \ref{fig:sinc_bper} displays a comparison between synthetic maps of the linearly polarized emission obtained for the  quasi-perpendicular case. These maps were obtained after considering different rotations with respect to the $x$- and $y$-axes.  It is important to note that these rotations only change the viewing angle, not the intrinsic geometry. The rotations with respect to the $x$- and $y$-axes, respectively, are indicated in the top right corner of each panel). Panel (a) displays a morphology which disagrees with the observed one \citep{Harvey-Smith2010} because it exhibits a strong central emission in addition to two bright and opposite arcs.  This strong central emission is mainly due to the $\mathrm{{\bf B}_{\varphi}}$ component of the field which has been swept up by the SNR shock wave. This emission is coming from the parts of the SNR shock front that are moving towards and away from us. The bright arcs are mainly due to the $\mathrm{{\bf B}_G}$ component which was compressed by the SNR shock wave.
Panels (b), (c), and (d) show only one bright arc, which is located on the right for panels (b) and (c), while for panel (d) this arc is on the left.

In Figure \ref{fig:sync_bpar} we compare the linearly  polarized emission for the quasi-parallel case, obtained with the same orientation as  Figure \ref{fig:sinc_bper}. In this case all panels show a clumpy structure. In panel (b), a bright arc is observed to the left, while panels (c) and (d) show two opposite bright arcs. Panel (c) displays a morphology that is closer to the observed one \citep{Harvey-Smith2010}. 

 In Figures 2 and 3 we observe that in the quasi-parallel case the angle of rotation along the $x$-axis determines the displacement in the vertical direction of the two bright arcs, while the angle of rotation along the $y$-axis determines the relative brightness of the two arcs. On the other hand, in the quasi-perpendicular case the angle of rotation along the $x$-axis determines the relative brightness of the two arcs, and the rotation angle along the $y$-axis has a very little effect on the synchrotron emission.

Figure \ref{fig:synchro} shows a comparison between intensities of the linearly polarized emission obtained from observations (panel a) and from numerical simulations for the quasi-parallel  (panel b) and quasi-perpendicular (panel c) cases.
%The observational image of the remnant $G296.5+10$ was obtained a a frequency of 1.4 GHz.
We obtained the synthetic polarized intensity maps from our numerical simulation by employing Eqs.(\ref{stokesQ})-(\ref{Ipol}). Comparing the synthetic maps (see Figures \ref{fig:sinc_bper} and \ref{fig:sync_bpar}) with the radio image at 1.4 GHz, we determined that the best results are obtained by rotating the computational domain by $40\degr$ around the $x$-axis and $5\degr$ around the $y$-axis for the quasi-parallel case, while for the quasi-perpendicular case the rotation were $-40\degr$ and $5\degr$ around $x-$ and $y-$ axes, respectively. As we can see the different particle injection models produce images that differ considerably from each other in appearance. A clear bilateral morphology with two bright arcs of emission to the left and right is obtained for the quasi-parallel case, in agreement with observations. Instead, the synthetic quasi-perpendicular map shows only a bright arc on the left.

\begin{figure}
   \centering
 \includegraphics[width=0.5\textwidth]{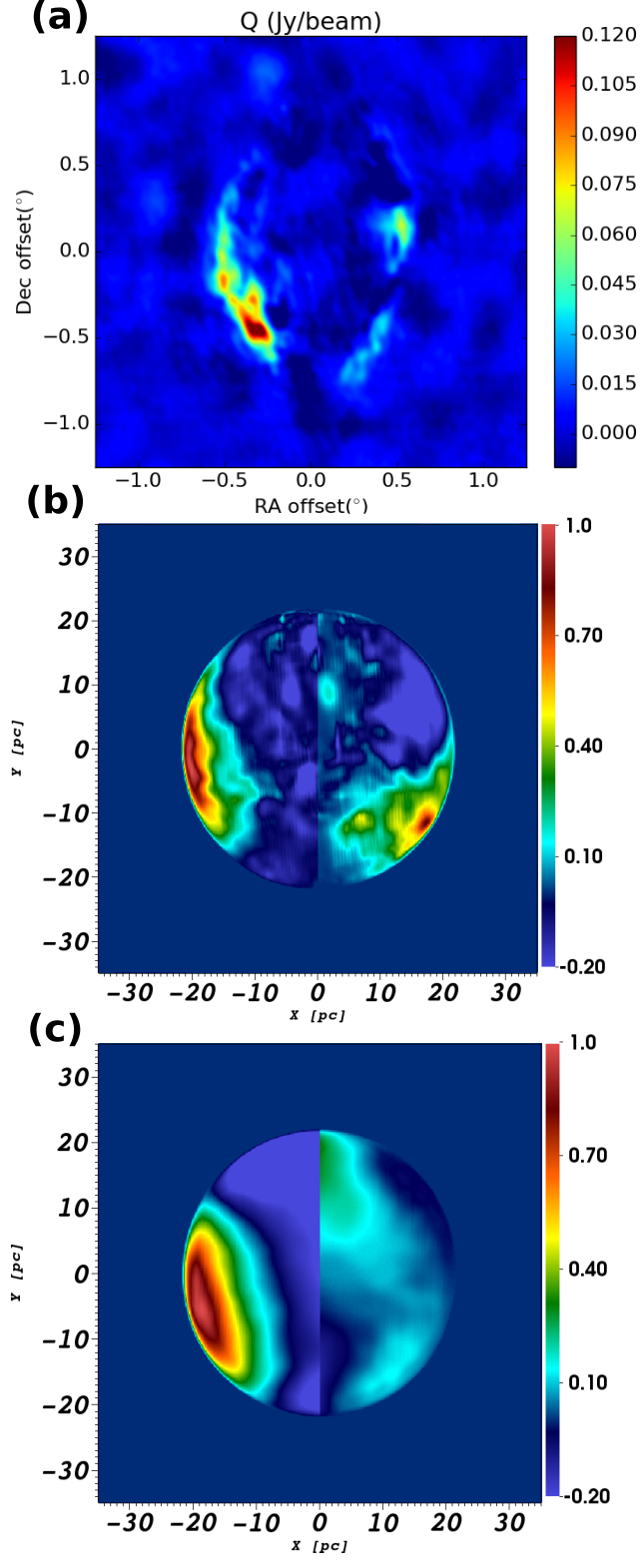}
\caption{Comparison between observations ({\it a}) and the synthetic Stokes parameters Q for the quasi-parallel ({\it b}) and quasi-perpendicular ({\it c}) cases. The axes are shown in parsecs.}
   \label{fig:Q}
\end{figure}

Figure \ref{fig:Q} shows a comparison of the Stokes parameter Q for the observed image (panel a) and the synthetic maps corresponding to the quasi-parallel (panel b) and quasi-perpendicular (panel c) cases. 
These maps were computed considering the  $\Delta\chi_F$ obtained from the RM values reported by \citet{Harvey-Smith2010}, which have opposite signs for the Eastern and Western sides of the remnant. As was mentioned in subsection 3.3, for the left half of the synthetic map we have used a $\Delta\chi_F=75.4\degr$, while a $\Delta\chi_F=-37.7\degr$ was employed for the right half of the simulated SNR. This produces the artificial left/right  discontinuity seen in panels (b) and (c). Both synthetic Q maps achieve a reasonable  agreement with the observational Q image.

\begin{figure}
   \centering
   \includegraphics[width=0.45\textwidth]{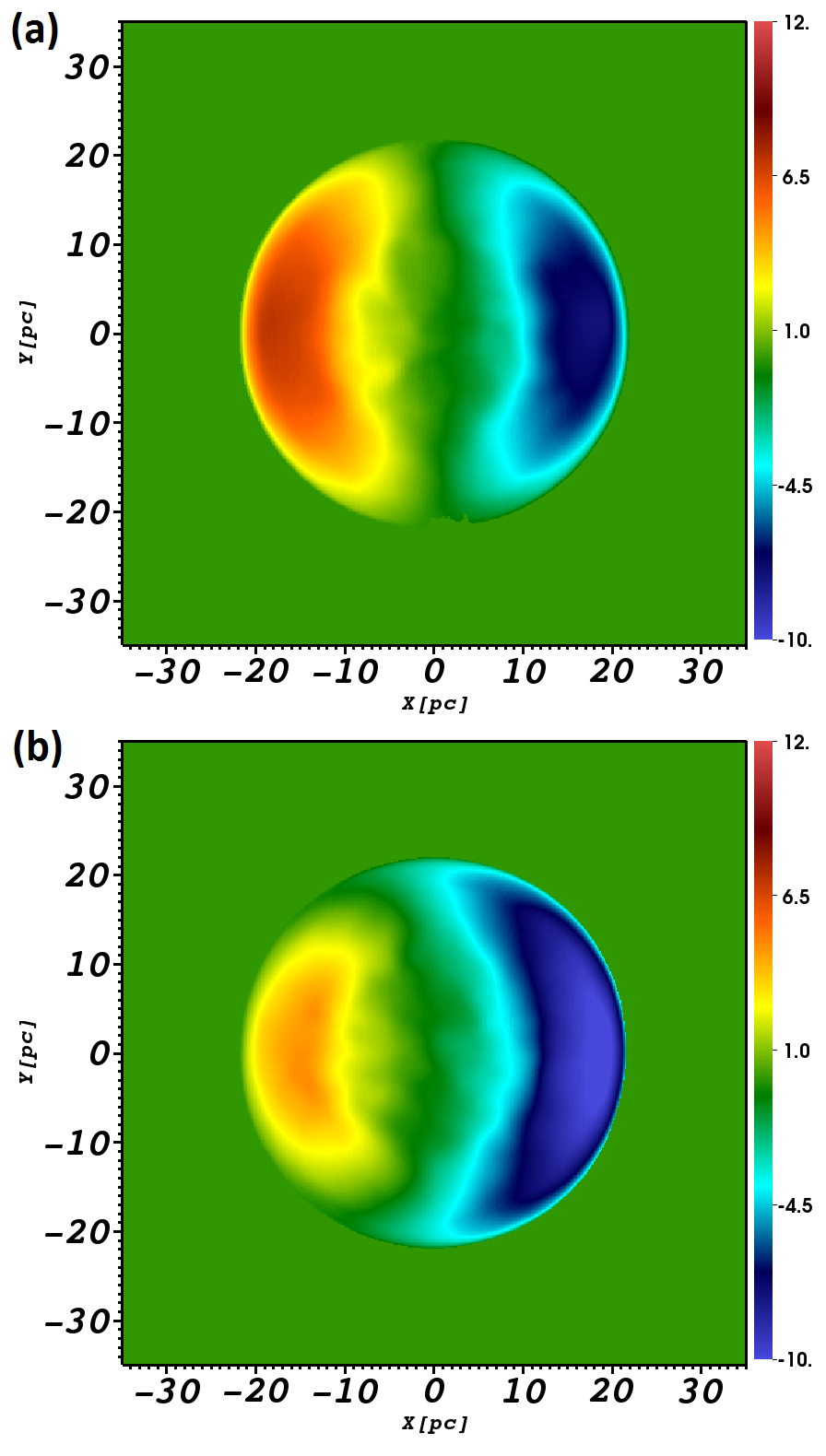}
   \caption{Rotation measure for the ({\it a}) quasi-parallel and ({\it b}) quasi-perpendicular cases. The color scale is given in units of $\mathrm{rad\ m^{-2}}$}
\label{fig:RM}
\end{figure}

Synthethic RM maps were also constructed for both cases and they are displayed in Figure \ref{fig:RM}. Both maps show a  positive RM in the East side and negative in the West side. However, the RM map obtained for the quasi-parallel case is in better agreement (taking into account the observational errors) with previous observational  works  \citep{Whiteoak-gardner68,Harvey-Smith2010}.
Note that these maps show the intrinsic RM. \citet{bandiera2016} have analyzed the influence of the intrinsic RM on both the Stokes parameter Q maps and the distribution of the magnetic field position angle. However, in our work the main objective is to do a direct comparison with the observations. Thus synthetic Stokes parameter Q maps must be corrected by the total RM, which is composed of the intrinsic RM and that due to $B_{\parallel}$ of the medium between the source and us. For this reason we employ the RM reported by observations \citep{Harvey-Smith2010} for the construction of the synthetic Stokes parameter Q maps shown in Figure \ref{fig:Q}.

\begin{figure}
   \centering \includegraphics[width=0.5\textwidth]{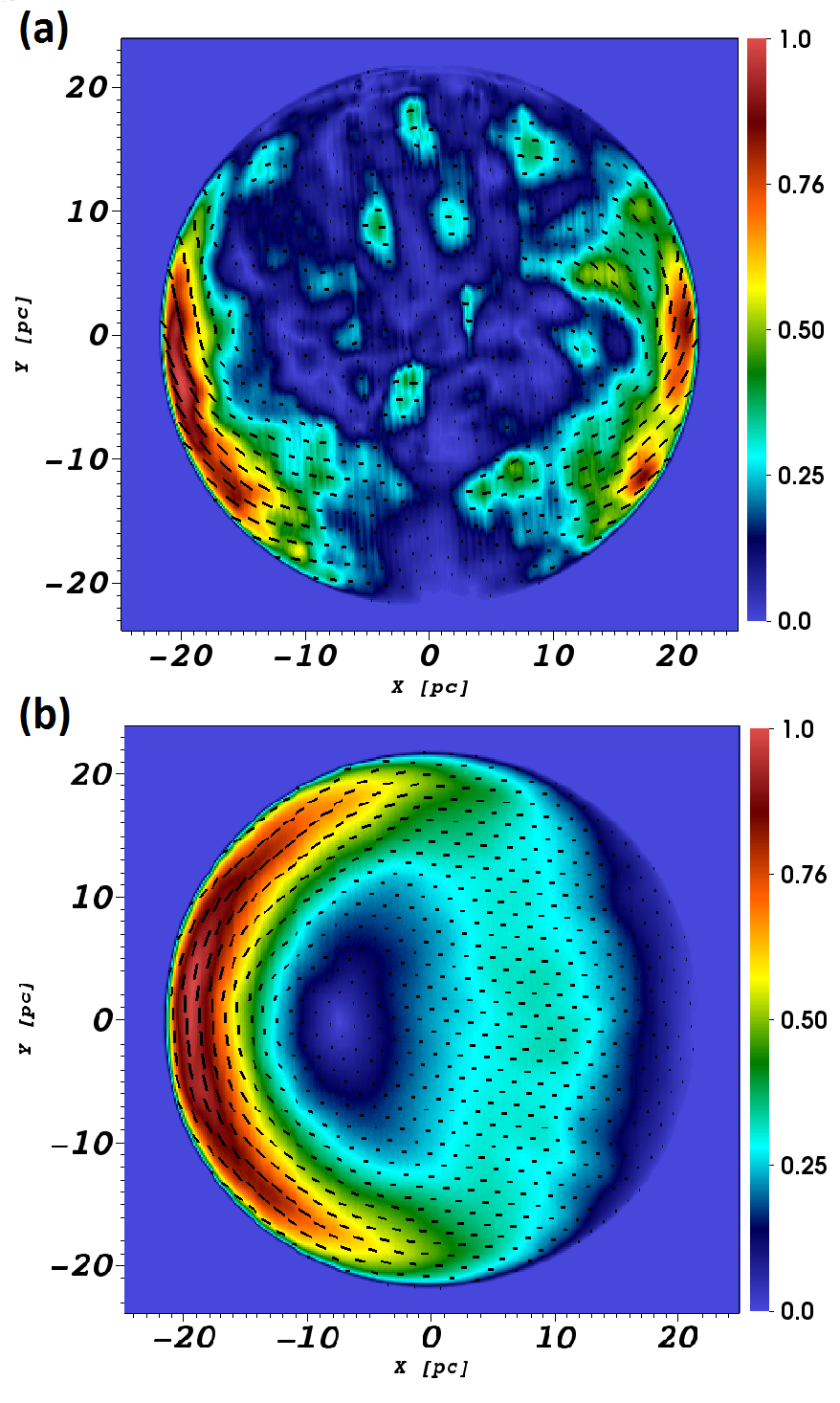}
\caption{Distribution of the intrinsic magnetic field (black lines) superimposed on the linearly polarized emission map (normalized to its maximum) for the (a) quasi-parallel and (b) quasi-perpendicular cases.}
   \label{fig:vec_B}
\end{figure}

Finally, we compute the intrinsic orientation of the projected magnetic vector on the plane of the sky for both the quasi-parallel and the quasi-perpendicular cases. To do this, we employed the position angle given by Equation (\ref{chi_B}) and oriented the map by following the conventional orientation, i.e. North pointing to up ($\hat{y}$) and East pointing to the left ($-\hat{x}$). Then, for each point on the plane of the sky,  we used the following equation to obtain the projected magnetic field:
\begin{equation}
\mathrm{\mathbf{B}}_{\text{pos}}(x_i,y_i) = -\sin(\chi_B(x_i,y_i)) \hat{x} + \cos(\chi_B(x_i,y_i)) \hat{y} \;.
\end{equation}
Figure \ref{fig:vec_B} displays the magnetic field distribution for the quasi-parallel and quasi-perpendicular cases (panels a and b, respectively). The black lines represent the orientation of the magnetic field with their lenght scaled with the linearly polarized intensity. We can see that the quasi-parallel case is in better agreement with observational results reported by \cite{Whiteoak-gardner68}, who found a tangential orientation of the intrinsic magnetic field around the bright edges of the SNR by analyzing $1.4$~GHz and $2.7$~GHz data. The same result was obtained by \citet{milne1994}, who presented intensity and polarization maps at 2.4, 4.8, and 8.4 GHz. It is important to note that for the quasi-parallel case model, there is not a significant radial component of the magnetic field (see Figure \ref{fig:vec_B}a), such as it was observed and reported for the SN 1006 \citep{reynoso2013,Schneiter15,velazquez2017}. The main reason for this is the fact that the azimuthal component from the progenitor's wind dominates the region of emission, after been swept up by the SNR shock wave.

\section{Discussion and Conclusions}

The synchrotron radio emission of the bilateral supernova remnant
G$296.5+10$ was compared with results obtained from MHD simulations. For the magnetic field we considered two components: (1) the magnetic field of the progenitor \citep[considering the hypothesis presented by][]{Harvey-Smith2010}, and (2) a uniform Galactic magnetic field. Two scenarios were considered for the acceleration mechanism of the relativistic particles: the quasi-parallel and the quasi-perpendicular one.

A polarization study of the synchrotron emission was carried out from the numerical results, from which we obtained synthetic maps of the linearly polarized emission, the Stokes parameter Q, the distribution of the intrinsic magnetic field projected in the plane of the sky and the rotation measure. These maps were compared with the corresponding observations at a frequency of 1.4 GHz.  As a result of this comparison we found that the quasi-parallel case is more adequate for reproducing the linearly polarized intensity, the synthetic RM maps \citep[yielding values similar to those reported by][]{Harvey-Smith2010}, and the magnetic field (projected on the plane of the sky) distribution, while the Q maps are very similar for both acceleration mechanisms.

It is important to note that the analysis of Q parameter maps shown to be a useful tool to discriminate the acceleration mechanism taking place in the SNR shock for the case of the SN 1006 \citep{Schneiter15}. For this remmant it was assumed  that this object expands into an ISM with an almost uniform magnetic field.

The present work however shows that the distribution of the Q parameter does not appear to be a determining factor in settling which acceleration mechanism is responsible for the linearly polarized emission. Our study rather shows that for a more complex magnetic field configuration (such as that used in the case of SNR G296.5+10), a joint analysis including the linearly polarized emission, the distribution of Q, and the magnetic field in the plane of the sky must be performed in order to discern between the quasi-parallel and the quasi-perpendicular mechanisms. 

We must highlight that our results differ from those reported by \citet{west2016} and \citet{orlando2007} for the case of this remnant. A possible explanation of this discrepancy could be attributed to the fact that these authors only considered the Galactic component of the magnetic field, while we also included the azimuthal component of the magnetic field of the progenitor. % Furthermore, they did not compare synthetic and observed maps of the Q parameter.

In summary, we found that the quasi-parallel mechanism produces a significantly better fit to the observed linearly polarized intensity, the Stokes parameter Q maps, the magnetic field distribution in the plane of the sky, and the RM values.
Our results are compatible with a scenario in which the SNR expands in a surrounding medium where the magnetic field is the sum of the uniform Galactic field and the azimuthal component of the magnetic field due to the progenitor star, as suggested by \citet{Harvey-Smith2010}.

\section*{Acknowledgements}

We thank Dr. Rino Bandiera (the referee) for his thorough reading of our manuscript, and his opportune suggestions and comments which help us to improve the previous version of this work. AM-B, PFV, AE, and FDC acknowledge finantial support from DGAPA-PAPIIT (UNAM) grants IG100516, IN109715, IA103315, and
IN117917. JCTR akcnowledges finantial support from DGAPA-PAPIIT (UNAM) grant IV100116. EG acknowledge support from ANPCYT PICT 2015-1729 and UBACYT 20020150100098BA funds.   AM-B is a fellow CONACyT (M\'exico). EG and EMS are  members of the Carrera del
Investigador Cient\'\i fico of CONICET, Argentina.
 We thank Enrique Palacios-Boneta (c\'omputo-ICN) for mantaining the Linux servers where our simulations were carried out.

\section*{Dedication}
%%Dedication%%
AM-B wants to dedicate this article to the memory of her father, Jorge A. Moranchel Castrej\'on. "Although today you are no longer with me, you will live forever in my memory, your teachings will always be the guide of my steps, and you will be the light of my path. I love you, and I will love you forever father"(AM-B).
%%%%%%%%%%%%%%%%%%%%%%%%%%%%%%%%%%%%%%%%%%%%%%%%%%



%\bibliographystyle{mnras}

% This method is tedious and prone to error if you have lots of
% references

\bibliography{ref}{}
\bibliographystyle{mnras}

% %%%%%%%%%%%%%%%%%%%%%%%%%%%%%%%%%%%%%%%%%%%%%%%%%%

% %%%%%%%%%%%%%%%%% APPENDICES %%%%%%%%%%%%%%%%%%%%%

% \appendix

% \section{Some extra material}

% If you want to present additional material which would interrupt the flow of the main paper,
% it can be placed in an Appendix which appears after the list of references.

%%%%%%%%%%%%%%%%%%%%%%%%%%%%%%%%%%%%%%%%%%%%%%%%%%

% Don't change these lines
\bsp	% typesetting comment
\label{lastpage}
\end{document}